\documentclass[3p,times]{elsarticle}
\usepackage{graphicx}
\usepackage{amsmath,amssymb}

\usepackage{lineno}

\begin{document}

\begin{frontmatter}



\title{Inclusive Jet Spectra in p-Pb Collisions at ALICE}

\author{Megan Connors (Yale University) for the ALICE Collaboration)}





\begin{abstract}

Jet suppression has been observed in central heavy ion
collisions. This suppression is attributed to partonic energy loss in
the Quark Gluon Plasma (QGP) formed in such collisions. However, this
measurement is influenced by all stages of the collision. It is
expected that in p-Pb collisions similar initial conditions occur as
in Pb-Pb collisions without creating a QGP, allowing modification to
the jet spectra due
to cold nuclear matter effects to be quantified. Inclusive jet spectra
in p-Pb collisions at
$\sqrt{s_{\rm NN}} = 5.02$ TeV measured by ALICE are presented.  Jets
are reconstructed via the anti-k$_{\rm T}$ algorithm with different
resolution parameters by combining charged tracks measured in the
ALICE tracking system with the neutral energy deposited in the
electromagnetic calorimeter. The jet spectra can be used to determine a nuclear modification factor $R_{\rm pPb}$ while the jet profile in p-Pb is studied by dividing spectra measured with different resolution parameters and comparing to the same ratio measured in pp collisions. 

\end{abstract}



\end{frontmatter}



\section{Introduction}
\label{intro}

Jets, the collimated sprays of particles resulting from hard
scattering processes, are an excellent probe for heavy ion
collisions. Since the hard scattering happens early in the collision,
the partons probe all stages of the collision while traversing the
produced medium. Suppression of jet production is quantified by comparing the fully
corrected jet spectrum measured in central Pb-Pb collisions to that measured in pp
scaled by the average number of binary collisions, $N_{\rm
  coll}$ \cite{alicePbPbchjets,alicePbPbjets}. This
suppression is attributed to partonic energy loss in the Quark Gluon
Plasma (QGP) created in heavy ion collisions. However, cold nuclear matter (CNM)
effects due to the initial state could also influence this
measurement. Measurements of the jet cross section in p-Pb collisions, which experience CNM effects without the creation of the final state
QGP are critical for disentangling initial and final state effects on
the observed Pb-Pb jet spectra.

\section{Analysis Details}
\label{analysis}

The present analysis uses minimum bias 5.02 TeV p-Pb events collected by the ALICE experiment
during the 2013 LHC run with an
integrated luminosity of 51 $\mu b^{-1}$. Input to the jet reconstruction
algorithm are
clusters, measured in the
electromagnetic calorimeter (EMCal) with $E_{T}>300$ MeV/$c$ and
charged tracks with $p_{T} > 150$ MeV/$c$ measured in the ALICE central tracking
system, which consists of a time projection chamber (TPC) and a silicon inner
tracking system (ITS). To correct the EMCal clusters for energy
deposited by charged tracks, 100\% of the momentum of any tracks
geometrically matching that cluster is subtracted from the cluster \cite{aliceppjets}.   
Jets are reconstructed with the anti-$k_{T}$ jet finding
algorithm with resolution parameters, $R$ = 0.2
and $R$ = 0.4 using the FastJet package \cite{fastjet}. To remove ``fake'' jets,
clustered constituents that did not originate from a hard scattering, the area of the jet,  $A_{jet}$, is required to
fulfill $A_{jet}>0.6\pi R^{2}$. To ensure the jet is fully within the
detector acceptance, the jet axis must be at least $R$ away from
the edge of the detector. The detector boundaries are defined by the EMCal acceptance, $|\eta|<0.7$ and $1.4< \phi < \pi$.

Energy from the underlying event is also clustered into the
jets by the algorithm and must be subtracted from the total raw jet
energy. An average energy density, $\rho$, is determined on an event-by-event
basis and then subtracted from each jet in the event according to
$p_{T,jet}^{reco}=p_{T,jet}^{raw}-\rho_{scaled} \times A_{jet}$. To
reduce the effect of the limited EMCal acceptance, we base the
background density on the charged track $p_{T}$ density, $\rho$, which
is measured in full azimuth and scale it up to $\rho_{scaled}$, using a scale factor that is
determined from measured data to include electromagnetic contributions, as done in the Pb-Pb jet spectra
analysis \cite{alicePbPbjets}. While for Pb-Pb the median method was used to
calculate the charged track background density, a median occupancy
method, which is a slightly modified implementation of that
presented in \cite{cmsrho}, is used here. The median occupancy method,
defined by
\begin{equation}
\label{eq:rho}
\rho=\rm{median}\left \{\frac{p_{T}^{i}}{A_{i}} \right \} \times C,
\end{equation}
is determined by running the $k_{T}$ algorithm over all tracks plus ``ghost
particles'' (unphysical particles with negligible momentum added
artificially by FastJet
for the jet area calculation) in the event \cite{fastjet}. Before determining the
median of the physical jets, any $k_{T}$ jets overlapping with signal jets are
excluded. Signal jets are defined as anti-k$_T$ jets with $p_{T} > 5$ GeV/$c$. The median is then scaled by an occupancy correction
factor, $C=A_{physical jets}/A_{all jet}$, where $A_{physical jet}$ is the total area of all physical jets and
$A_{all jet}$ is the area of all jets, including jets that only
contain ghost particles. This factor, $C$, accounts for the emptiness of
the p-Pb event. 



\subsection{Unfolding}
\label{uf}

\begin{figure}[htp]
\centerline
{
\includegraphics[width=0.5\textwidth]{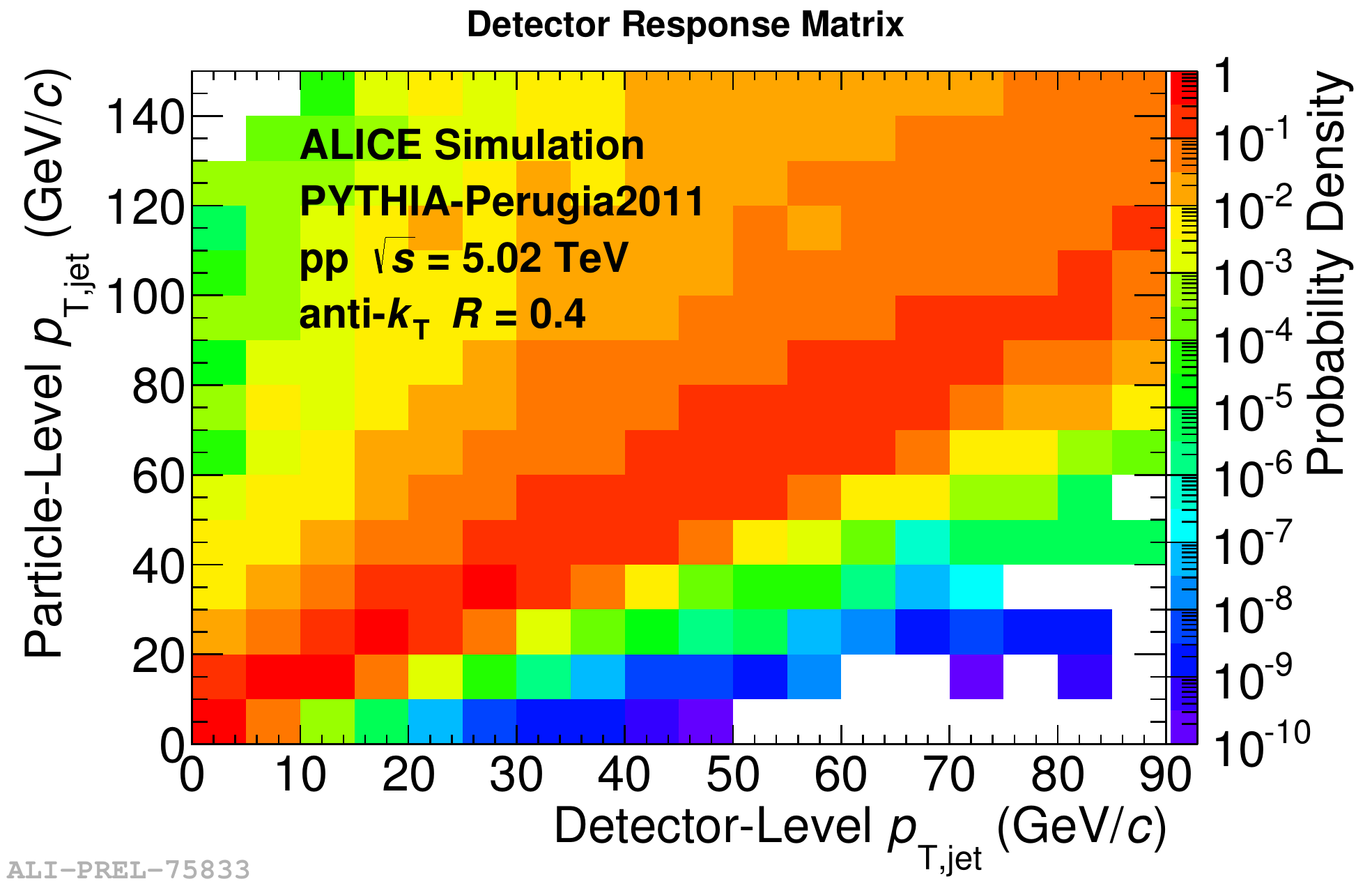}
\includegraphics[width=0.42\textwidth]{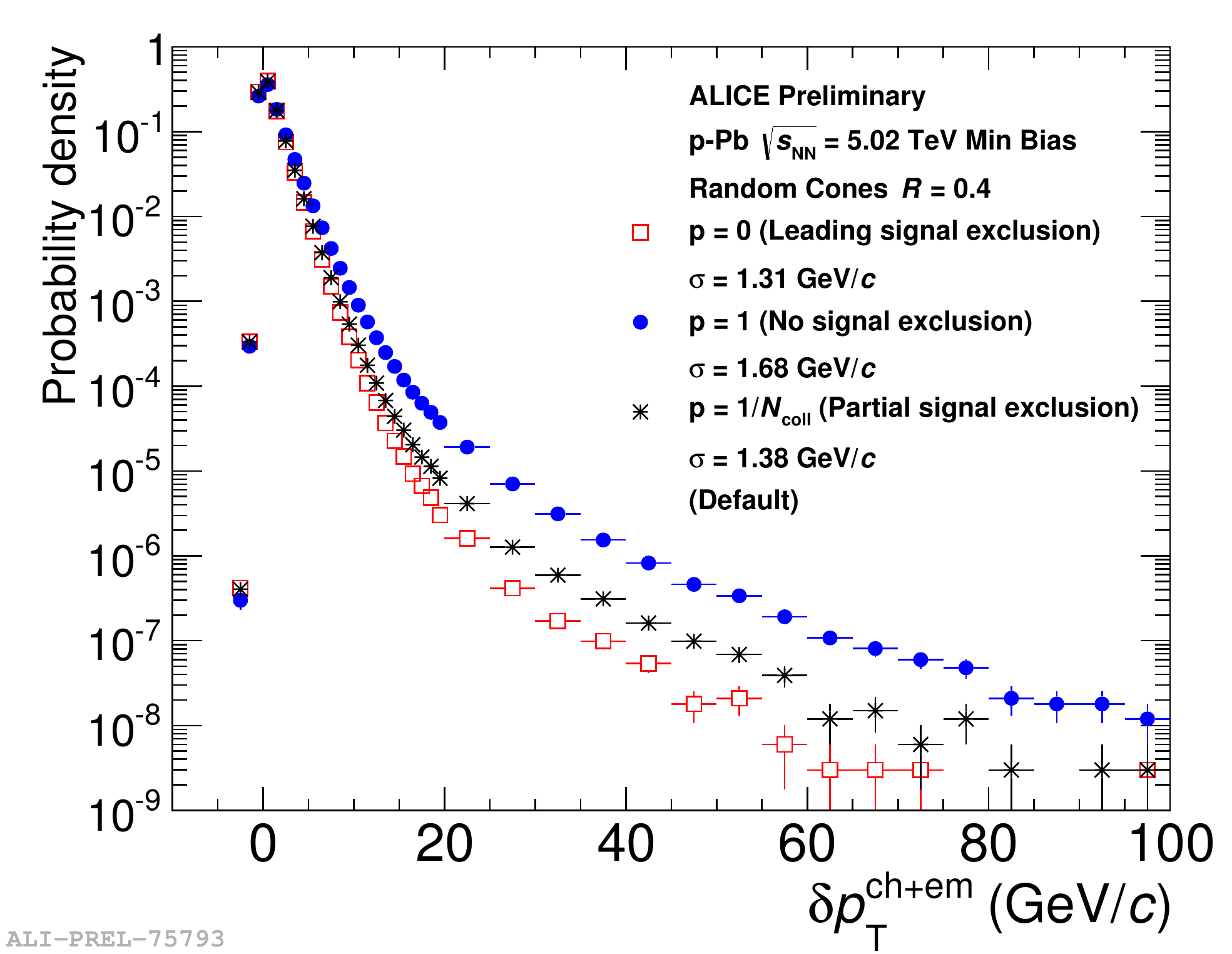}
  }
  \caption{(Left) Detector response matrix mapping the reconstructed detector level jet $p_{T}$ to
    particle level jet $p_{T}$. (Right) The $\delta p^{ch+em}_{T}$ determined
    using Random Cones. The leading jet from each event were
    removed with a
    probability of $1/N_{\rm coll}$ (black astrics) for the analysis. To estimate the
    systematic uncertainty,
    all (red open squares) and no leading jets are removed (blue circles). Both plots are for anti-k$_{T}$ jets
    with $R$ = 0.4.}
\label{fig:RMdpt}
\end{figure}

The measured spectra must be corrected for detector effects and the
influence of fluctuations on the underlying event. The effect of the detector on the spectra is determined by
passing PYTHIA events at $\sqrt{s} =$ 5.02 TeV through a GEANT
simulation of the ALICE detector. Jets are reconstructed at the
detector level using the same cuts as applied on the data. Jets
are also reconstructed at the particle level without cuts to
give the true jet $p_{T}$. All
detector level jets are geometrically matched to the nearest particle
level jet. A
2-dimensional histogram or response matrix (RM) maps
detector level jet $p_{T}$ to particle level jet $p_{T}$ and is used in the
unfolding procedure. Figure \ref{fig:RMdpt} shows an example of the detector
RM for $R$ = 0.4. A correction
is also applied to account for the jet reconstruction efficiency. This correction is determined by dividing the true
PYTHIA jet spectrum by a projection of the
RM onto the particle level jet $p_{T}$ axis.

The underlying event energy density is determined on an event-by-event
basis, however, even within a single event  there are fluctuations
within the background energy density.  The momentum of the jet may be over-
or underestimated if it was positioned on an upward
or downward fluctuation. These fluctuations can be quantified by
measuring the $\delta p^{ch+em}_{T}$ distribution using the method of
Random Cones (RC) according to
\begin{equation}
\delta p^{ch+em}_{T}=p_{T}^{RC}-\pi R_{RC}^2 \times \rho, 
\end{equation}
where the $\rho$ of the
event is compared to the total momentum, $p_{T}^{RC}$, within a cone of
radius, $R_{RC}$, placed randomly in the event.  The right panel of Figure
\ref{fig:RMdpt} shows the $\delta p^{ch+em}_{T}$ distribution for $R_{RC}$ = 0.4. The long tail on the right
hand side of the distribution represents the probability of having two
overlapping jets. Since the random cone can be thought of as introducing an
additional jet to the event, this probability is artificially enhanced. We account for this effect by
excluding the leading jet from the event at a rate of  $1/N_{\rm coll}$ (black
points). To estimate the systematic uncertainty on this procedure,
this exclusion probability can be varied between zero, where no jets are removed, and one,
where all leading jets are removed. The $\delta p^{ch+em}_{T}$ distribution with no jets removed and all
leading jets removed from each event are shown as blue circles and
red squares respectively in Figure \ref{fig:RMdpt} and result in a  
3\% uncertainty on the final spectra.

The final RM, the multiplication of the detector RM and the $\delta p^{ch+em}_{T}$ distribution, is input to the unfolding algorithm. The singular value decomposition (SVD) algorithm was chosen as the
default for unfolding the spectrum \cite{svd}. Unfolding with the Bayesian method was also performed
and the difference used in the estimation of the systematic
uncertainties \cite{bayes}. A bin-by-bin correction procedure
was found to be in good agreement with the other methods. 

\begin{figure}[htp]
\centerline
{
\includegraphics[width=0.42\textwidth]{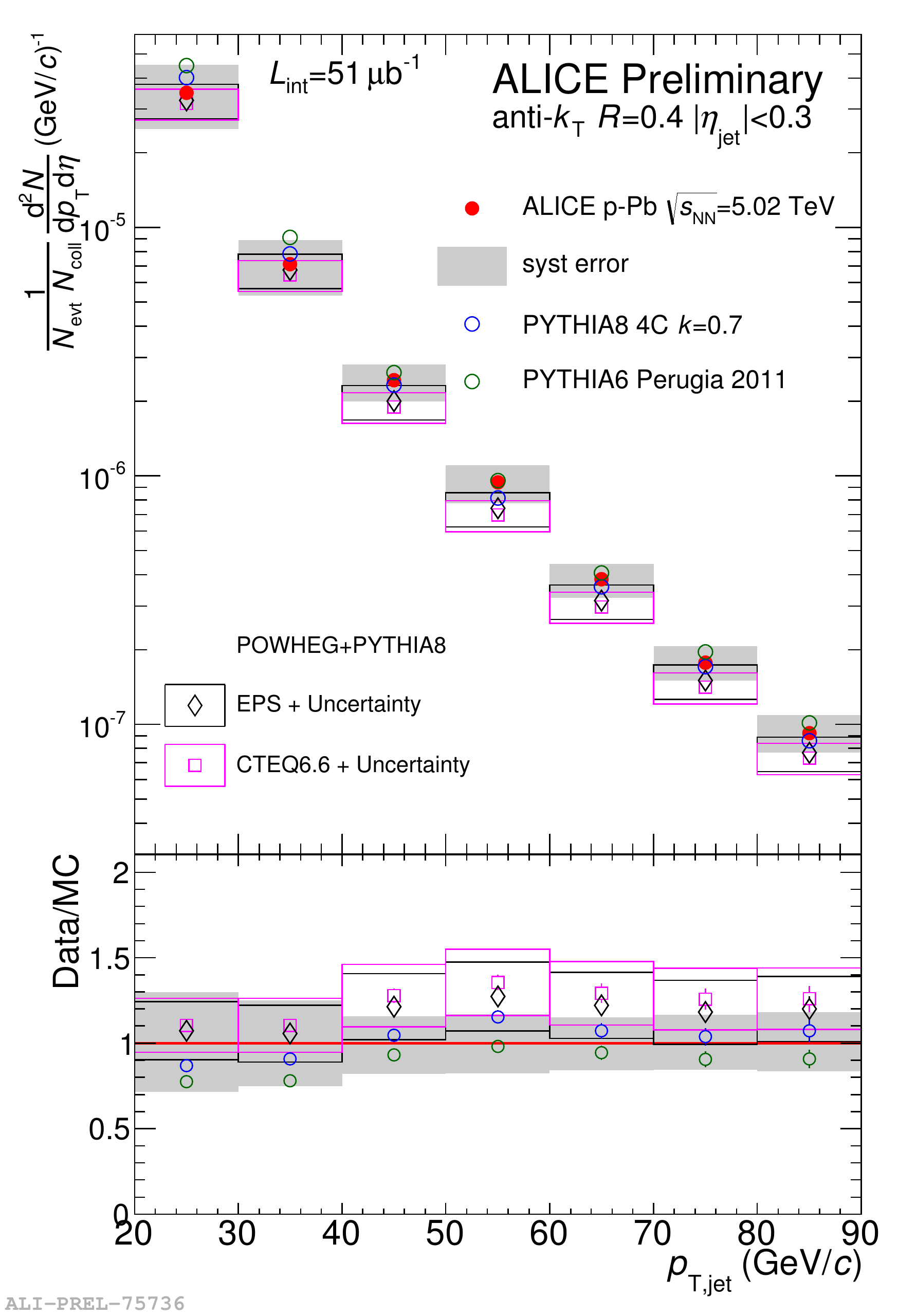}
\includegraphics[width=0.42\textwidth]{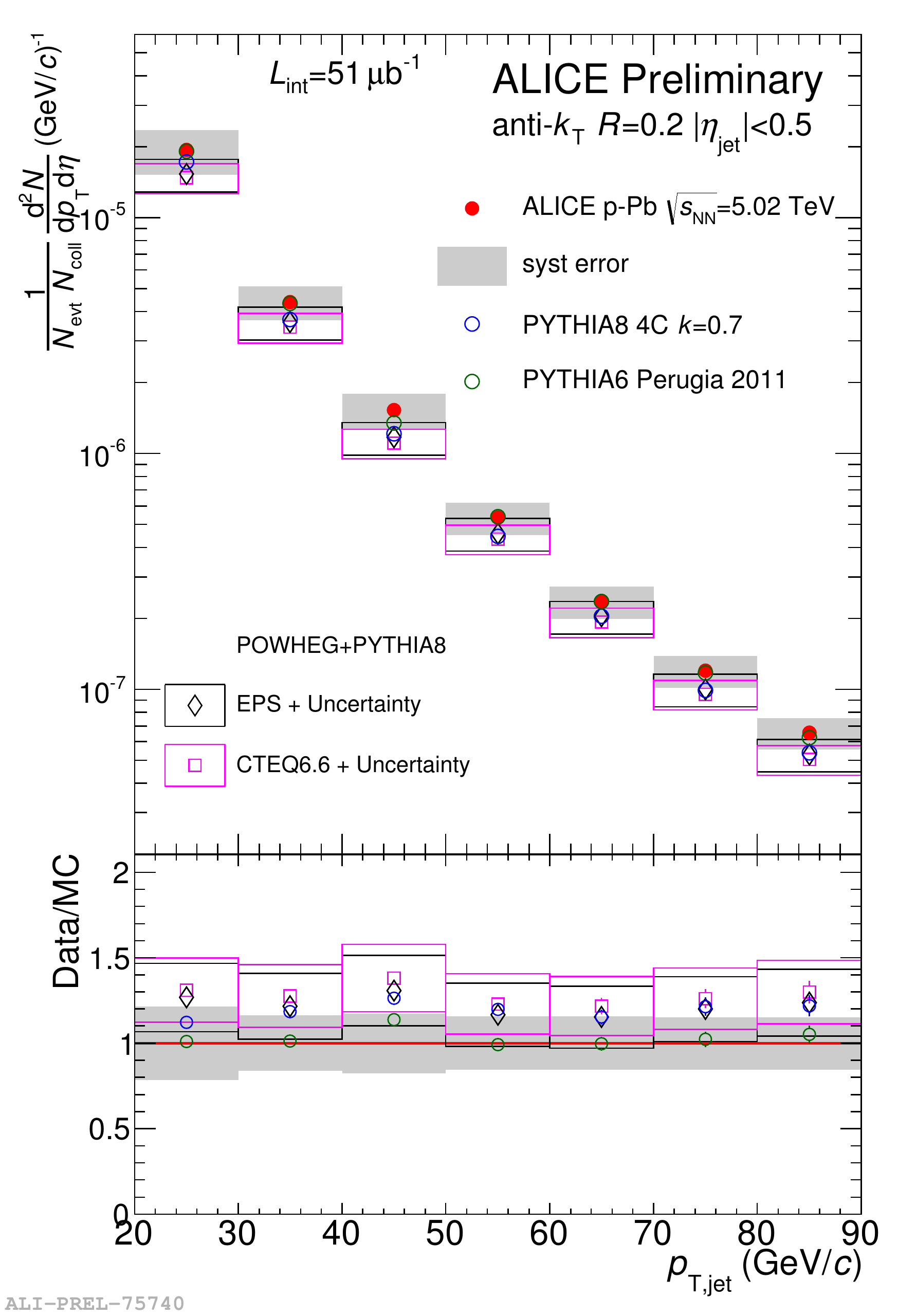}
  }
  \caption{Fully corrected p-Pb jet spectrum for $R$ = 0.4 and $R$ = 0.2
    scaled by $N_{\rm coll}$
    compared to PYTHIA and POWHEG simulations at 5.02 TeV.}
\label{fig:spectra}
\end{figure}
\section{Results}

The resulting spectra after unfolding for $R$ = 0.4 and $R$ = 0.2 are shown in
Figure \ref{fig:spectra}. The spectra were normalized per average number of
binary collisions, $N_{\rm coll}$, to make a direct comparison to the pp
references. Due to the lack of a measured pp reference at
$\sqrt{s}$ = 5.02 TeV, we compare the p-Pb data to simulations. The plot
includes PYTHIA8, PYTHA6 and POWHEG with 2 different parton
distribution functions (PDF). The POWHEG calculations include
uncertainties on the factorization and renormalization scales (13\%) and the uncertainty in the PDF
(6\% for CTEQ and 9\% for EPS). The ratios between the data and the
different models are all consistent with one, indicating there are no
cold nuclear matter effects to the jet spectrum. However, the spread and uncertainty from these
different references is significant and highlights the need for a data
reference to better quantify this statement. Despite this uncertainty,
the p-Pb results clearly demonstrate that the strong suppression observed in the Pb-Pb is not
purely due to initial state effects, but is rather a result of
energy loss in the produced medium. 

\begin{figure}[htp]
\centerline
{
\includegraphics[width=0.6\textwidth]{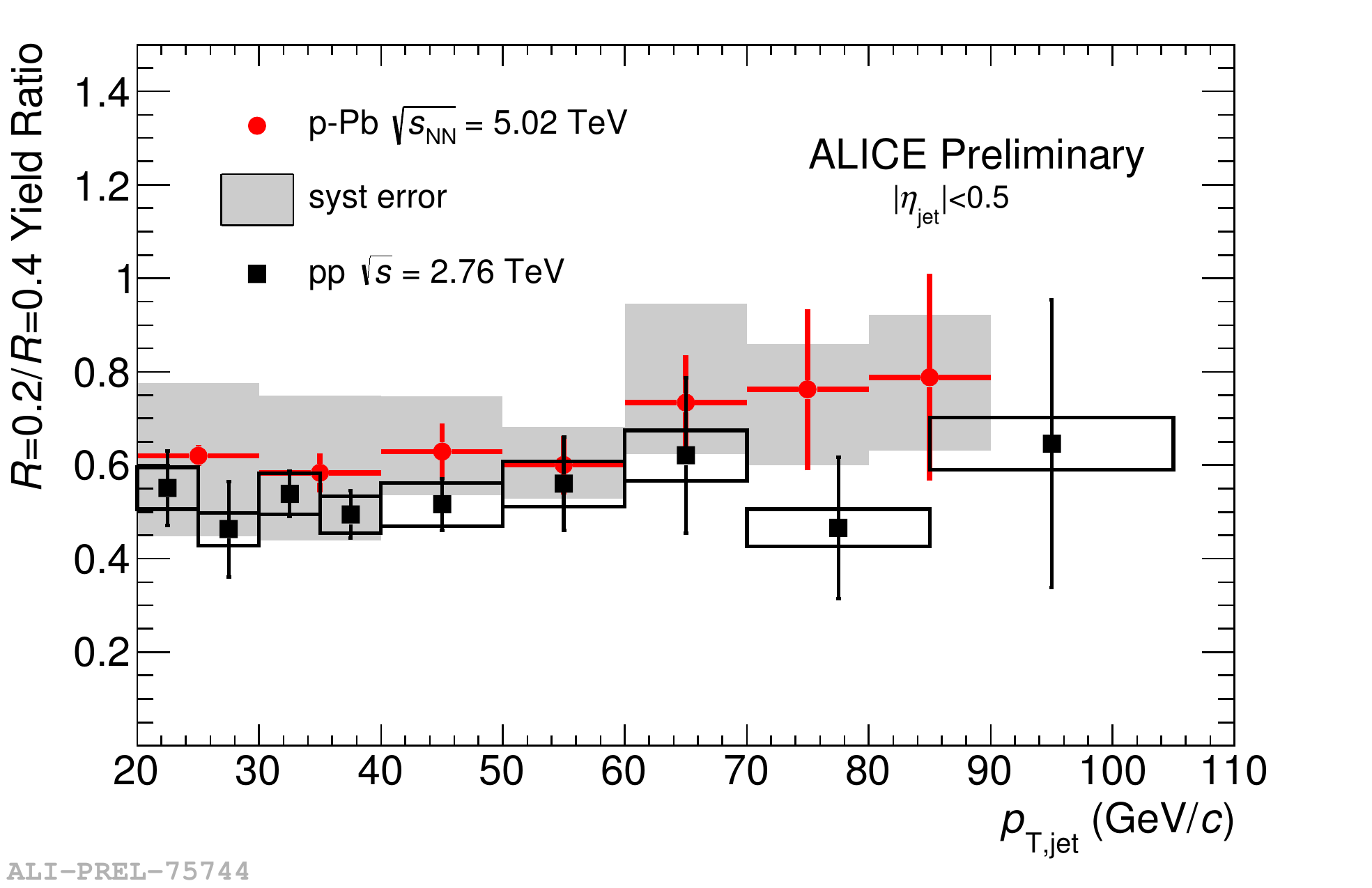}
  }
  \caption{Cross-section ratio between $R$ = 0.2 jet spectrum and $R$ =
    0.4
    jet spectrum for fully reconstructed jets in 5.02 TeV p-Pb
    collisions (red) and 2.76 TeV pp collisions (black).}
\label{fig:JSR}
\end{figure}

One may question whether the fragmentation could be modified in p-Pb
collisions while the total jet cross section is not. A first
indication of 
the fragmentation behavior of these jets can be obtained by taking
the ratio of the spectra measured with different $R$. Figure
\ref{fig:JSR} shows the ratio of the $R$ = 0.2 spectrum to the $R$ = 0.4
spectrum for 5.02 TeV p-Pb (red circles) and for 2.76 TeV pp (black
squares) collisions.  The agreement between the two collision systems
suggests that the fragmentation behavior for jets in p-Pb is very
similar to that in pp collisions. Of course, more differential studies
will provide more detailed insight into the fragmentation properties.




\section{Conclusions}

The fully reconstructed jet spectra for $R$ = 0.2 and $R$ = 0.4 have been
measured by ALICE in 5.02 TeV p-Pb collisions in the $p_{T}$ range 20-90 GeV/$c$.  Comparisons to model
predictions of the 5.02 TeV pp jet spectra indicate that the strong
suppression observed in Pb-Pb collisions is an effect of the medium and
not an initial state effect. To better quantify the CNM effects, if
any, on the jet spectrum, systematic uncertainties on this measurement must be reduced. In
particular, the uncertainty on the reference can be reduced by measuring
pp collisions at 5 TeV at the LHC. The ratio between spectra
reconstructed with different $R$ is consistent with the same ratio in pp
collisions. This also indicates no modification to the substructure of the
jet. Inclusion of EMCal triggered events will extend the
kinematic reach of this measurement and allow for multiplicity
dependent studies.








\end{document}